\documentclass[twocolumn,journal,comsoc]{IEEEtran}

\newtheorem{theorem}{Theorem}{}
{}
{}

\usepackage[T1]{fontenc}

\usepackage{amsmath,amsfonts}
\usepackage[cmintegrals]{newtxmath}
\usepackage{color,bm}
\usepackage{url,epsfig,graphicx}
\usepackage[caption=false,font=footnotesize]{subfig}
\usepackage{algorithm,algpseudocode}
\usepackage{multirow}

\usepackage[dvipdfm,bookmarksopen,colorlinks,linkcolor=black,anchorcolor=black,citecolor=black]{hyperref}

\newcommand{\field}[1]{\mathbb{#1}}
\newcommand{\C}{\field{C}}
\newcommand{\R}{\field{R}}

\newcommand{\abs}[1]{\left\lvert#1\right\rvert}
\newcommand{\norm}[1]{\left\lVert#1\right\rVert}
\newcommand{\CT}{\text{H}}
\newcommand{\T}{\text{T}}
\newcommand{\st}{\text{s.t.} \quad }
\newcommand{\vect}[1]{\bm{#1}}
\newcommand{\mat}[1]{\bm{#1}}

\DeclareMathOperator{\Real}{Re}
\DeclareMathOperator{\Imag}{Im}
\DeclareMathOperator{\Tr}{Tr}
\DeclareMathOperator{\argmin}{argmin}
\DeclareMathOperator{\argmax}{argmax}

\begin{document}


\title{Joint Multicast Beamforming and User Scheduling in Large-scale Antenna Systems}

\author{Longfei~Zhou,~\IEEEmembership{Student Member,~IEEE,} Zi~Xu,
	Wei~Jiang,~\IEEEmembership{Member,~IEEE,}
	and~Wu~Luo,~\IEEEmembership{Member,~IEEE}
	
	\thanks{This work was supported
		by the National Natural Science Foundation of China under grant 61171080. }
	\thanks{L. Zhou, W. Jiang, and W. Luo are with the State Key Laboratory of Advanced
		Optical Communication Systems and Networks, Peking University, 100871, China. 
		E-mail: \{zhoulongfei,jiangwei, luow\}@pku.edu.cn.}
	\thanks{Z. Xu are with Department of Mathematics, College of Sciences, Shanghai University, 
		Shanghai 200444, People's Republic of China. E-mail: xuzi@i.shu.edu.cn} 
}

\maketitle

\begin{abstract}
This paper studies the joint multicast beamforming and user scheduling problem, with the objective of minimizing total 
transmitting power across multiple channels by jointly assigning each user to appropriate channel
and designing multicast beamformer for each channel. 
The problem of interest is formulated in two
different optimization problems, a mixed binary quadratically constrained quadratic program and
a highly-structured nonsmooth program.
Two different algorithms, based on convex relaxation and
convex restriction, respectively, are proposed to solve the problem. The performance ratio between 
the approximate solution provided by the convex-relaxation-based algorithm 
and optimal solution is proved to be upper bounded by a constant independent of problem data.
The convex-restriction-based algorithm is guaranteed to converge to a critical point to the 
nonsmooth formulation problem. Finally, extensive simulation results verify the theoretical 
analysis and demonstrate the advantage of the proposed co-design scheme over conventional 
fixed scheduling and random scheduling in terms of power consumption.
\end{abstract}


\IEEEpeerreviewmaketitle

\section{Introduction}\label{sec1}
Demands for high-rate wireless services, such as Internet TV, on-line gaming, 
and multimedia downloading, continue to grow explosively in the worldwide.
Wireless multicast is regarded as one of key enabling technologies in future 
cellular systems to boost the capacity of wireless networks and cater to the customer demands.
When combined with large-scale antenna arrays at base station (BS), wireless multicast is 
able to take full advantage of available channel state information at transmitter (CSIT) to 
provide enhanced data rates and relatively high transmission reliability \cite{2013SPM-MMIMO}.
Wireless multicast has always been an important part of the evolution of multimedia 
broadcast multicast service (MBMS) in wireless communication standards such as UMTS, LTE and LTE-Advanced   \cite{2012CM-eMBMS}.

Lots of downlink multicast beamforming problems have been discussed for different scenarios.
Single-group multicast beamforming for single-cell system was first investigated in \cite{2006TSP-SDR},
and then extended to multi-group multicast in \cite{2008TSP-MultiGroup}.
Multicast beamforming with per-antenna power constraints was further
discussed in \cite{2014-multicastPAPC} \cite{2015-multicastPAPC}.
Furthermore, coordinated multicast beamforming under per-BS power constraints for
multi-cell system was considered in \cite{2013-coordinated} \cite{zhou2017coordinated}.
Some other issues, such as energy efficient design, user selection and real-time implementation, were also studied in
\cite{2015-EnergyEfficientMulticast} \cite{2015-coordinatedMulticastingUserSelection} \cite{zhou2016fast}.

A commonly-used formulation of above studies is transmitting power minimization under quality of 
service (QoS) constraints. A key difficulty with such formulation is that the problem may be 
infeasible, especially when the number of users is much larger than the number of antennas. 
In such a situation, part of users should be removed out (admission control) or scheduled in orthogonal
resource dimensions, such as time, frequency, and code slots, which is  crucial for practical applications. 
The former leads to a variety of joint beamforming and admission control problems.

In \cite{matskani2009efficient}, the authors addressed the joint multicast beamforming and admission control 
problem based on semidefinite programming relaxation (SDR) and greedy membership deflation.
The basic idea is sequentially dropping a weakest user, then solving the relaxed problems 
and finally checking whether the suboptimal rank-one solution satisfies all QoS constraints.  
Recently, network energy efficient design and sparse optimization of the joint multicast beamforming and 
admission control for green Cloud-RAN was further discussed in \cite{shi2016smoothed}.
For the particular satellite communication systems, system sum rate optimization of the
multi-group multicast precoding and user scheduling under per-antenna power constraints 
and underlying framing structure constraints was considered in \cite{christopoulos2015multicast}.
In \cite{zhou2015joint}, a closed-form asymptotically optimal solution was proposed for the joint 
multi-group multicast beamforming and user grouping in massive MIMO systems.

This paper studies the joint multicast beamforming and user scheduling in large-scale 
antenna systems with a massive number of users. 
We assume that each user takes interest in multiple information symbols
but is assigned to receive one of its interested information symbols. Each information symbol
is transmitted over an orthogonal channel, such as, time slot and frequency subcarrier. 
The problem of interest is minimizing the total transmitting power across all orthogonal channels 
by jointly assigning each user to appropriate channel and designing multicast beamforming 
vector for each information symbol such that each user should successfully decode  
at least one information symbol. Since channel quality of each user at all channels should be 
taken into consideration, this problem is much different from the admission control 
in \cite{matskani2009efficient}, where only channel quality of all users at a fixed channel
is considered.

Our main contributions are summarized as follows. First, the problem of interest is cleverly formulated in two
different optimization problems, a mixed binary quadratically constrained quadratic program and
a highly-structured nonsmooth program. Second, a polynomial-time SDR algorithm is proposed to address
the problem. The worst-case approximation ratio of SDR is proved to be $\mathcal{O}(QK)$ for general channel scenario,
and $\mathcal{O}(K^{1/Q})$ for the special case of homogeneous channel scenario, where $Q$ and $K$ is the number
of orthogonal channels and the number of users, respectively. Our result is an important improvement 
and generalization upon those in \cite{wu2012rank} \cite{xu2014semidefinite} \cite{xu2016semidefinite}. 
Third, a sequential convex approximation (SCA) scheme 
is proposed for the nonsmooth formulation problem and an efficient dual fast gradient projection (DFGP) algorithm 
is devised for the subproblems. The overall algorithm is matrix-free, i.e., based solely on matrix-vector
multiplications and comparison operations, and guaranteed to converge to a critical point to the 
nonsmooth formulation problem. Finally, extensive simulation results are provided to verify the theoretical 
analysis and demonstrate the advantage of the proposed co-design scheme over conventional 
fixed scheduling and random scheduling in terms of transmitting power consumption.

The remainder of paper is outlined as follows. 
Section \ref{PF} describes the system model and the  two problem formulations. 
Section \ref{SDRA} presents the SDR algorithm and theoretical performance analysis.
Section \ref{SCA-DFGP} details the SCA-DFGP algorithm and its convergence result and computational complexity.
Section \ref{SR} and Section \ref{Conc} provides comprehensive simulation results to 
assess the performance of the proposed algorithms and concludes the paper, respectively.

\textit{Notation}: In the rest of this paper, boldface italic lowercase and 
uppercase characters denote column vectors and matrices, respectively. 
The operators $(\cdot)^\T, (\cdot)^\CT, \abs{\cdot}, \Tr(\cdot), \norm{\cdot}_2,$
and $\norm{\cdot}_F,$ correspond to the transpose, the conjugate transpose, the
absolute value, the trace and the Euclidean norm and the Frobenius norm operations,
while $\Real(\cdot)$ and $\Imag(\cdot)$ denotes the real part and 
imaginary part of complex number, respectively.

\section{System Model and Problem Formulation}\label{PF}
\subsection{System Model}
We consider a downlink multicast scenario consisting of a BS 
with $M$ antennas and $K$ single-antenna users.
Assume that there are $Q$ orthogonal channels $\mathcal{C}_q\:(q\in\mathcal{Q},\mathcal{Q}=\{1,2,\dots,Q\})$
between the BS and each user, such as nonoverlapping time slots or orthogonal subcarriers.
Let $\vect{\tilde{h}}_{k,q} \in \C^{M}$ denote the
complex channel vector between the BS and the $k$-{th} user for channel $\mathcal{C}_q$. 
Note that for each user these $Q$ channel vectors could be identical if the 
coherence bandwidth or coherence time is sufficiently large. 
Such a special case will be referred to as homogeneous channel scenario.
The BS uses an $M\times1$ beamforming vector $\vect{w}_q$ 
to send a zero-mean and unit-variance common information symbol $x_q$ to the interested users  
over channel $\mathcal{C}_q$. The  signal received from $\mathcal{C}_q$ by the $k$-{th} user is 
\begin{equation}
y_{k,q} = \vect{\tilde{h}}_{k,q}^\CT\vect{w}_q x_q + n_{k,q} \: \forall q\in\mathcal{Q}\: \forall k\in \mathcal{K},
\end{equation}
where $\mathcal{K}=\{1,2,\dots,K\}$ is the user index set, and $n_{k,q}$ is 
the zero-mean  circularly-symmetric complex Gaussian random noise with variance $\sigma_{k,q}^2$, 
which is independent of $x_q$ and $\vect{\tilde{h}}_{k,q}$.
The signal-to-noise ratio (SNR) at the $k$-{th} user can be expressed as
\begin{equation}
\gamma_{k,q} = \frac{\abs{\vect{\tilde{h}}_{k,q}^\CT\vect{w}_q}^2}{\sigma_{k,q}^2}\: \forall q\in\mathcal{Q}\:\forall k\in \mathcal{K}.
\end{equation}

The QoS requirement for the $k$-{th} user to successfully decode information symbol $x_q$
can be expressed as $\gamma_{k,q} \ge \bar{\gamma}_q.$ 
Let $\vect{h}_{k,q} = \vect{\tilde{h}}_{k,q}/(\sigma_{k,q}\sqrt{\bar{\gamma}_q})$
be the  $k$-{th} user's normalized channel vector for $\mathcal{C}_q \: (q\in\mathcal{Q})$.
The QoS requirement can be rewritten as
\begin{equation}\label{QoS_req}
\abs{\vect{h}_{k,q}^\CT\vect{w}_q}^2 \ge 1.
\end{equation}

We assume that each user takes interest in multiple information symbols
but is assigned to receive one of its interested information symbols.
When there are a large number of users in the system, it is impractical or inefficient to 
serve all users within a single channel. Therefore, properly scheduling all users to multiple channels
is important to boost the system capacity. 
\subsection{MBQCQP Formulation}
A commonly-used disjunctive modelling technique is 
using binary variable $b_{k,q}\in \{0,1\}$ as scheduling indicator,
i.e., $b_{k,q}=1$ indicates that the $k$-{th} user is scheduled in channel $\mathcal{C}_q.$
Hence, the problem of interest can be formulated as the following mixed binary quadratically 
constrained quadratic program (MBQCQP) 
\begin{subequations}\label{MBQCQP}
	\begin{align}
	\min_{\{\vect{w}_q\},\{b_{k,q}\}} &\quad \sum_{q=1}^{Q}\norm{\vect{w}_q}_2^2 \\
	\st & \abs{\vect{h}_{k,q}^\CT\vect{w}_q}^2 \ge b_{k,q} \: \forall q\in\mathcal{Q}\:\forall k \in \mathcal{K}, \\
	& \sum_{q=1}^{Q}  b_{k,q} = 1 \: \forall k \in \mathcal{K},  \\
	& b_{k,q} \in \{0,1\} \: \forall q\in\mathcal{Q}\:\forall k \in \mathcal{K}.
	\end{align}
\end{subequations}
In the above problem, the objective function are quadratic in the continuous variables 
and the disjunctive constraints contain both continuous and binary variables.
This class of MBQCQP problem is extremely difficult partly as they are nonconvex 
even with the binary variables being fixed \cite{xu2014semidefinite} \cite{xu2016semidefinite}. 
In the special case of $b_{k,1}=1$ for all 
$k\in \mathcal{K},$ the problem reduces to the single-group multicast beamforming problem, 
which is a continuous QCQP and NP-hard in general  \cite{2006TSP-SDR}.

\subsection{Nonsmooth Reformulation}
For each user, ensuring the QoS requirement \eqref{QoS_req} in at least one channel
is equivalent to making the QoS requirement in the best channel be satisfied.
Hence, the feasible set of continuous variables $\{\vect{w}_q\}$ in \eqref{MBQCQP} can 
be equivalently described by the following nonsmooth constraints
\begin{equation}
\max_{q\in\mathcal{Q}}\left\{\abs{\vect{h}_{k,q}^\CT\vect{w}_q}^2\right\} \ge 1 \: \forall k \in \mathcal{K}. 
\end{equation}
Denote $\mat{W}=[\vect{w}_1,\vect{w}_2,\dots,\vect{w}_Q] \in \C^{M\times Q}$ and
\begin{equation}\label{f_k}
f_k(\mat{W})=\max_{q\in\mathcal{Q}}\left\{\abs{\vect{h}_{k,q}^\CT\vect{w}_q}^2\right\}.
\end{equation}
Observing that  the binary variables in \eqref{MBQCQP} is absent from the objective function,
we obtain a nonsmooth reformulation of \eqref{MBQCQP} as follows
\begin{subequations}\label{NLQCQP}
	\begin{align}
	\min_{\mat{W} \in \C^{M\times Q}}& \quad \norm{\mat{W}}_F^2  \\
	\st & f_k(\mat{W}) \ge 1 \: \forall k \in \mathcal{K}.\label{nonsmooth_constrt}
	\end{align}
\end{subequations}

In this equivalent reformulation,  all binary variables are removed out at the expense of  
a small number of nonsmooth constraints. The main obstacle in \eqref{NLQCQP} is, of course, 
the nonsmoothness and nonconvexity of constraints. 
However, each constraint function $f_k(\mat{W})$ is highly structured,
and making use of the available structure in an appropriate way will give efficient algorithms
to solve \eqref{NLQCQP}. After solving \eqref{NLQCQP}, we can properly assign each user to 
the channel in which the user attains the best QoS among all channels.

We will propose a semidefinite relaxation (SDR) approach in section \ref{SDRA} to solve \eqref{MBQCQP},
and an efficient nonsmooth optimization approach in section \ref{SCA-DFGP} to solve \eqref{NLQCQP}.

\section{Semidefinite Relaxation Approach}\label{SDRA}
In this section, a SDR technique with performance guarantee is developed for solving the MBQCQP formulation \eqref{MBQCQP}. 
The main idea is to simultaneously use the continuous relaxation for the binary variables 
and the SDR for the continuous variables. After solving the SDR problem, a randomization procedure is used to 
generate approximate solutions to the original MBQCQP formulation from an optimal solution of the SDR problem. 
Furthermore, we analyze the bound on the approximation ratio between the optimal value 
of the  MBQCQP problem and that of the associated SDR.

Upon changing the optimization variables to $\mat{W}_q=\vect{w}_q\vect{w}_q^\CT$ and 
then doing the SDP relaxation for $\mat{W}_q$ and the continuous relaxation for $\{b_{k,q}\}$ 
in \eqref{MBQCQP}, we obtain the following problem
\begin{subequations}\label{SDR}
	\begin{align}
	\min_{\{\mat{W}_q\},\{b_{k,q}\}} &\quad \sum_{q=1}^{Q}\Tr(\mat{W}_q)\\
	\st & \vect{h}_{k,q}^\CT\mat{W}_q\vect{h}_{k,q}\ge b_{k,q} \: \forall q\in\mathcal{Q}\:\forall k \in \mathcal{K},  \\
	& \sum_{q=1}^{Q}  b_{k,q} = 1 \: \forall k \in \mathcal{K},  \\
	&  0\le b_{k,q} \le 1 \: \forall q\in\mathcal{Q}\:\forall k \in \mathcal{K}, \\
	& \mat{W}_q \succeq 0 \: \forall q\in\mathcal{Q}.
	\end{align}
\end{subequations}
We observe that these continuous variables $\{b_{k,q}\}$ in $\eqref{SDR}$ can be eliminated out from the problem without
loss of optimality. An equivalent problem is obtained as follows
\begin{subequations}\label{SDR1}
\begin{align}
	\min_{\{\mat{W}_q\}} &\quad \sum_{q=1}^{Q}\Tr(\mat{W}_q) \\
	\st & \sum_{q=1}^{Q}\vect{h}_{k,q}^\CT\mat{W}_q\vect{h}_{k,q} \ge 1 \: \forall k \in \mathcal{K}, \\
	& \mat{W}_q \succeq 0 \: \forall q\in\mathcal{Q}.
\end{align}
\end{subequations}
One can verify that each feasible solution to \eqref{SDR1} is also feasible to \eqref{SDR} 
and vice versa. Moreover, the same formulation is obtained if similar convex 
relaxation is applied to the nonsmooth problem \eqref{NLQCQP}.

For the special case of homogeneous channel scenario that the $Q$ channel vectors 
for each user are identical, 
i.e., $\vect{\tilde{h}}_{k,1}=\vect{\tilde{h}}_{k,2}=\dots=\vect{\tilde{h}}_{k,Q}\stackrel{\Delta}{=}\vect{\tilde{h}}_{k} \:\forall q \in \mathcal{Q}$, problem \eqref{SDR1} is symmetric with respect to the arguments $\{\mat{W}_q\}$ and could be further reduced to
\begin{subequations}\label{SDR2}
	\begin{align}
    \min_{\mat{W}_1} &\quad \Tr(\mat{W}_1)  \\
	\st & \vect{\tilde{h}}_{k}^\CT\mat{W}_1\vect{\tilde{h}}_{k} \ge 1 \: \forall k \in \mathcal{K}, \\
	& \mat{W}_1 \succeq 0.
	\end{align}
\end{subequations}
Surprisingly, besides much simpler formulation, it will been shown in next subsection that 
\eqref{SDR2} provides better performance guarantee for homogeneous channel scenario 
than for general channel scenario.

Problem \eqref{SDR1} and \eqref{SDR2} are both convex, which can be efficiently solved using 
off-the-shelf interior point solvers such as SDPT3 and SeDuMi.
Once an optimal solution $\{\mat{W}_q^*\}$ to \eqref{SDR1} is obtained
[for \eqref{SDR2}, $\mat{W}_1^*=\mat{W}_2^*=\dots=\mat{W}_Q^*$], the Gaussian randomization method could be used 
to generate the candidate beamformers. The $l$-th candidate beamformer for 
channel block $\mathcal{C}_q$ is generated
as $\vect{x}_q^{(l)}=\mat{U}_q\mat{\Sigma}_q^{1/2}\vect{v}_l,$ where $\mat{U}_q,\mat{\Sigma}_q$ 
are the eigen-decomposition factors of $\mat{W}_q^*,$ i.e., $\mat{W}_q^*=\mat{U}_q\mat{\Sigma}_q\mat{U}_q^\CT,$ 
and  $\vect{v}_l\sim\mathcal{CN}(\vect{0},\mat{I}_M).$ It's easy to show 
that $\vect{x}_q\sim\mathcal{CN}(\vect{0},\mat{W}_q^*).$
Given such candidate beamformers $\{\vect{x}_q\}_{q=1}^Q$, we still need to determine 
the corresponding transmitting power. Let $c_{k,q} = \abs{\vect{h}_{k,q}^\CT\vect{x}_q}^2$. 
Substituting $\vect{w}_q=\sqrt{p_q} \vect{x}_q$ into \eqref{NLQCQP}, we have
\begin{subequations}\label{PA}
	\begin{align}
	&\min_{\vect{p}\ge 0} \quad \sum_{q=1}^{Q}p_q\norm{\vect{x}_q}^2\\
	\st & \max_{q\in\mathcal{Q}}\left\{p_q c_{k,q}\right\} \ge 1 \: \forall k \in \mathcal{K}. \label{constrt}
	\end{align}
\end{subequations}

Albeit nonconvex, the problem \eqref{PA} is a special monotonic optimization problem.
Many kinds of outer approximation algorithms could be applied\cite{zhang2013monotonic}. 
For the sake of analysis convenience, a simple scaling procedure is used to obtain high-quality approximate solution. 
The feasible approximate solution to \eqref{PA} is given by 
$p_q=p(\{\vect{x}_q\}) \: \forall q\in\mathcal{Q},$ where 
\begin{equation}\label{p_x}
p(\{\vect{x}_q\})=  \frac{1}{\min_{k \in \mathcal{K}}\max_{q\in\mathcal{Q}} \{c_{k,q}\}}.
\end{equation}

For completeness, the overall SDR-G algorithm for \eqref{MBQCQP} or \eqref{NLQCQP} is summarized in Algorithm 1.
\begin{algorithm}[!t]
	\caption{SDR-G algorithm for \eqref{MBQCQP} or \eqref{NLQCQP} }
	\begin{algorithmic}[0]
		\State  \textbf{Initialization} Solve an optimal solution $\{\mat{W}_q^*\}$ to \eqref{SDR1}. 
		\For{$l=1,\dots,L$}
		\State  \textbf{1)} Sample $\vect{x}_q^{(l)}\sim\mathcal{CN}(\vect{0},\mat{W}_q^*)\:(\forall q\in\mathcal{Q}).$ \\
		\State \textbf{2)} Calculate $p^{(l)} = p(\{\vect{x}_q^{(l)}\})\sum_{q\in\mathcal{Q}}\norm{\vect{x}_q^{(l)}}^2$ using \eqref{p_x}. 
		\EndFor
		\State  \textbf{Output} Let $l^*=\argmin_{l=1,\dots,L} p^{(l)}.$ Select $\{\sqrt{p^{(l^*)}}\vect{x}_q^{(l^*)}\}$ as the approximate solution to \eqref{NLQCQP}.
	\end{algorithmic} 
\end{algorithm}

\subsection{Approximation Ratio}
In this subsection, we will analyze the performance of proposed SDR-G algorithm.
Denote the optimal value of SDR problem \eqref{SDR1} by $v_{\text{SDR-LB}}^*$, the optimal value of MBQCQP 
problem \eqref{MBQCQP} by $v^*$, and the objective value of the approximate solution to \eqref{NLQCQP} 
by $v_{\text{SDR-G}}^*$. Obviously, we have 
\begin{equation}\label{sandwich1}
v_{\text{SDR-LB}} \le v^* \le v_{\text{SDR-G}}^*=\min_{l=1,\dots,L} p^{(l)}. 
\end{equation}
We will show that there exists a constant $\theta > 0$ only depending on the number of orthogonal channels $Q$
and the number of users $K,$ such that 
\begin{equation}\label{sandwich2}
v_{\text{SDR-G}}^* \le \theta v_{\text{SDR-LB}}
\end{equation}
holds true with overwhelming probability. Such a constant $\theta$ is generally referred 
to as approximation ratio in computational complexity theory. It implies that the power loss 
due to the SDR approximation is at most $10\log_{10} \theta \text{\:dB}$  
away from the optimal transmitting power $v^*$ according  to \eqref{sandwich1} and \eqref{sandwich2}.
The main results about the upper bound on the worst-case approximation ratio $\theta$ are 
given in the following theorem.

\begin{theorem}
	(1) For general channel scenario,
	\begin{equation}
	\theta \le 5QK
	\end{equation}
	holds with probability at least $1-0.9^L.$ \\
	(2) For the special case of homogeneous channel scenario,
	\begin{equation}
	\theta \le 5K^{1/Q}
	\end{equation}
	holds with probability at least $1-0.9^L.$	
\end{theorem}
Please refer to Appendix for the proof of Theorem 1. Let's give some physical meaning explanations about why 
the worst-case approximation ratio are different between two scenarios. For general channel scenario,
when $Q-1$ channels are very poor simultaneously for all users, then scheduling all users into the 
rest channel are optimal. This degenerated problem is nothing but single-group multicast problem,
for which the worst-case performance bound of $\mathcal{O}(K)$ provided by SDR is in fact tight up 
to a constant factor \cite{luo2007approximation}. For homogeneous channel scenario,
the bound of $\mathcal{O}(K^{1/Q})$ can be regarded as the result of a kind of user selection 
diversity according to the proof. For average-case general channel scenario, one could expect such 
diversity, which, however, vanishes in the worst-case scenario. For the special case of homogeneous 
channel scenario with $Q=2$, a bound of $\mathcal{O}(K)$ is shown in \cite{xu2016semidefinite}. 
Moreover, by using a rank-two transmit beamformed Alamouti space-time code scheme for single-group multicast,
a bound of $\mathcal{O}(\sqrt{K})$ is obtained in \cite{wu2012rank}.
Our result is an interesting improvement and generalization upon above results. 

\section{Nonsmooth Optimization Approach}\label{SCA-DFGP}
Although SDR is a valuable benchmark for the problem, the computational burden of SDR is not 
well scalable to large-scale antenna system. Moreover, the worst-case results imply that 
the performance of SDR-G may deteriorate considerably when there are a massive number of users in the system.
Hence, we also provide an efficient algorithm to handle with such case.
MBQCQP formulation \eqref{MBQCQP} is difficult to solve due to a great number of binary variables and 
disjunctive constraints. We turn to highly-structured 
nonsmooth problem \eqref{NLQCQP}. Specifically, we devise a sequential approximation scheme to 
yield a series of smooth convex subproblems, and present a dual fast gradient projection algorithm 
to solve each subproblem. Finally, convergence and computational complexity of the overall algorithm is analyzed.

\subsection{Sequential Convex Approximations}
Since nondifferential constraint function $f_k(\mat{W})$ in \eqref{f_k} is the maximum of 
a finite number of convex quadratic functions, $f_k(\mat{W})$ is convex as well. Therefore, 
we have the following subgradient inequality,
\begin{equation}\label{subgrad_ineq}
f_k(\mat{W}) \ge f_k(\mat{V}) + \langle \mat{G}_{k}(\mat{V}),\mat{W}-\mat{V}\rangle \: \forall \mat{W},
\end{equation}
where $\mat{G}_{k}(\mat{V})$ is a subgradient of $f_k(\mat{W})$ at $\mat{V}$ and
$\langle\mat{G},\mat{W} \rangle  = \Real\left(\Tr(\mat{G}^\CT\mat{W})\right)$ 
is the inner product of two complex matrices $\mat{G}$ and $\mat{W}.$ 

At differentiable points, there is a unique subgradient of $f_k(\mat{W}),$ i.e., the 
gradient, while at nondifferentiable points, there is an infinite set of subgradients.
All subgradients $\mat{G}_{k}(\mat{W})$ of $f_k(\mat{W})$ satisfying \eqref{subgrad_ineq} 
form a convex set called subdifferential.
According to subdifferential calculus of convex functions \cite{boyd2004convex}, 
we can write the subdifferential of $f_k(\mat{W})$ as
\begin{equation}
\begin{split}
\partial f_k(\mat{W}) =  &\left\{  \left[\vect{d}_1,\vect{d}_2,\dots,\vect{d}_Q \right] : \right. \\
&\vect{d}_q = 2\alpha_q\vect{h}_{k,q}\vect{h}_{k,q}^\CT\vect{w}_q \:\forall q\in\mathcal{Q}, \\
& \alpha_q = 0 \: \forall q \notin I_k(\mat{W}), \\
& \sum_{q \in I_k(\mat{W})} \alpha_q = 1,  \alpha_q \ge 0 \left. \right\}, \\
I_k(\mat{W}) =&  \left\{q : \abs{\vect{h}_{k,q}^\CT\vect{w}_q}^2 = f_k(\mat{W})\right\}.
\end{split}
\end{equation}
Our choice of subgradient is 
\begin{equation}\label{sparseSub}
\mat{G}_{k}(\mat{W}) = \frac{2}{\abs{I_k(\mat{W})}}\sum_{q \in I_k(\mat{W})}\vect{h}_{k,q}\vect{h}_{k,q}^\CT\vect{w}_q,
\end{equation}
where $\abs{I_k(\mat{W})}$ is the cardinality of set $I_k(\mat{W})$.
The idea behind such choice is ensuring equal probability of scheduling each user into the active channels.

By iteratively linearizing $f_k(\mat{W})$ at $\mat{W}^{(n)}$ using \eqref{subgrad_ineq} and \eqref{sparseSub}, 
we obtain a sequence of convex approximations of problem \eqref{NLQCQP} as follows
\begin{subequations}\label{SCA}
	\begin{align}
	& \min_{\mat{W} \in \C^{M\times Q}} \quad \norm{\mat{W}}_F^2 \\
	\st & \langle\mat{G}_{k}(\mat{W}^{(n)}),\mat{W} \rangle + c_k(\mat{W}^{(n)}) \ge 0,\forall k \in \mathcal{K}.
	\end{align}
\end{subequations}
where $c_k(\mat{W}^{(n)}) = f_k(\mat{W}^{(n)})-\langle\mat{G}_{k}(\mat{W}^{(n)}),\mat{W}^{(n)} \rangle-1.$

Problem \eqref{SCA} is a strongly-convex quadratic program and therefore has a unique solution. 
Since the convergence rate of the subgradient-based method for nonsmooth optimization problem 
may be slow, efficient subproblem-solving algorithm is necessarily important, 
which will be detailed in next subsection.

\subsection{Dual Fast Gradient Projection Method}\label{DFGPA}
Since the constraints in \eqref{SCA}  are all linear inequalities 
and $\mat{W}^{(n)}$ is a feasible solution to \eqref{SCA},
the refined Slater's condition for \eqref{SCA} is satisfied \cite{boyd2004convex}. 
It implies that strong duality holds, i.e.,
the  optimal  value of \eqref{SCA} is equal  to  the  attained optimal  value  of the  dual problem.
Due to strong convexity of problem \eqref{SCA}, its dual problem is Lipschitz smooth and could 
be solved efficiently by a fast gradient projection method. 

To avoid complex notations, we consider the following general model of problem \eqref{SCA}
\begin{subequations}\label{CQP}
	\begin{align}
	\min_{\vect{x} \in \C^{MQ}}  & \quad \norm{\vect{x}}_2^2 \\
	\st & \Real(\mat{A}\vect{x}) + \vect{a} \ge 0.
	\end{align}
\end{subequations}
where $\mat{A}\in \C^{K \times (MQ)}$ and $\vect{a}\in \R^{K}.$ Obviously, problem \eqref{SCA}
could be cast into \eqref{CQP} by appropriate matrix concatenation.

The Lagrangian function associated with \eqref{CQP} is 
\begin{equation}
L(\vect{x},\vect{z}) = \vect{x}^\CT\vect{x} - \Real(\vect{z}^\T\mat{A}\vect{x}) - \vect{z}^\T\vect{a}.
\end{equation}
Minimizing $L(\vect{x},\vect{z})$ over  $\vect{x}$ gives 
the optimal solution
\begin{equation}\label{opt_sol}
\vect{x}=\frac{1}{2}\mat{A}^\CT\vect{z}
\end{equation}
and the dual objective function 
\begin{equation}
D(\vect{z}) = - \frac{1}{4}\vect{z}^\T\mat{B}\vect{z} - \vect{a}^\T\vect{z},
\end{equation}
where $\mat{B} =\Real(\mat{A}\mat{A}^\CT) = \Real(\mat{A})\Real(\mat{A})^\T + \Imag(\mat{A})\Imag(\mat{A})^\T.$

Hence, the dual program of \eqref{CQP} has a same solution set with the following problem
\begin{subequations}\label{DCQP}
	\begin{align}
	\min_{\vect{z}\in \R^{K}} & \frac{1}{4}\vect{z}^\T\mat{B}\vect{z}+\vect{a}^\T\vect{z} \\
	\st & \vect{z} \ge 0.
	\end{align}
\end{subequations}
Problem \eqref{DCQP} is a continuously-differentiable convex
minimization problem with a very simple constraint set.
Applying Nesterov's optimal gradient scheme \cite{2013-nesterov-introductory} to \eqref{DCQP}, we obtain 
the dual fast gradient projection (DFGP) iteration formula as follows
\begin{subequations}\label{DFGP}
	\begin{align}
	\vect{z}^{(l)} & =\max\left(\vect{\tilde{z}}^{(l-1)}-\mu(\frac{1}{2}\mat{B}\vect{\tilde{z}}^{(l-1)}+\vect{a}),0\right), \\
	\vect{\tilde{z}}^{(l)} & = \vect{z}^{(l)} +\frac{l-1}{l+2}\left(\vect{z}^{(l)}-\vect{z}^{(l-1)}\right),
	\end{align}
\end{subequations}
where $\mu=\frac{2}{\lambda_1(\mat{B})}$ and $\lambda_1(\mat{B})$ is the maximum eigenvalue 
of positive semidefinite matrix $\mat{B}.$ It is known that the algorithm converges to an $\varepsilon$-optimal
solution to \eqref{DCQP} within $\mathcal{O}(\frac{1}{\sqrt{\mu\varepsilon}})$ iterations \cite{2013-nesterov-introductory}. 

\subsection{Convergence and Complexity}
For clarity, the overall algorithm for the nonsmooth reformulation \eqref{NLQCQP} is summarized in Algorithm 2.
\begin{algorithm}
	\caption{SCA-DFGP algorithm for \eqref{NLQCQP}}  
	\textbf{output}: $\mat{W}^{(n)} $
	\begin{algorithmic}
		\State \textbf{Initialization} Randomly generate a feasible initial point $\mat{W}^{(0)}.$
		\For{$n=1,2,\dots$}
		\State \textbf{Step 1.} Calculate $\mat{G}_{k}(\mat{W}^{(n-1)})$ according to \eqref{sparseSub} for all $k.$
		\State \textbf{Step 2.} Solve \eqref{SCA} for the solution $\mat{W}^{(n)}$  using \eqref{DFGP} and \eqref{opt_sol}.
		\State \textbf{Step 3.} \textbf{If} $\norm{\mat{W}^{(n)}-\mat{W}^{(n-1)}}_F \le \varepsilon$, \textbf{then} STOP. 
		\EndFor
	\end{algorithmic} 
\end{algorithm} 
We first analyze the convergence of proposed algorithm. Let $\mathcal{P}(\mat{W}^{(n)})$ 
denote the instance of problem \eqref{SCA} at $\mat{W}^{(n)}.$ Since the cost function $\norm{\mat{W}}_F^2$
 is independent of $n$ and $\mat{W}^{(n)}$ is also feasible for $\mathcal{P}(\mat{W}^{(n+1)}),$ we have strict inequality $\norm{\mat{W}^{(n+1)}}_F^2 < \norm{\mat{W}^{(n)}}_F^2$ unless $\mat{W}^{(n+1)}=\mat{W}^{(n)}.$ Hence, the cost sequence $\left\{\norm{\mat{W}^{(n)}}_F^2\right\}$ converges either in finite iterations or to a unique value. 
Noting the cost function is exactly the square of Frobenius norm of $\mat{W}$, the variable sequence $\left\{\mat{W}^{(n)}\right\}$  converges to a unique point $\mat{W}^{(\infty)}$ as well. 
Since the feasible set in $\eqref{NLQCQP}$ is semi-algebraic, $\mathcal{P}(\mat{W}^{(n)})$ is an inner 
convex approximation of $\eqref{NLQCQP}$, and each constraint function of $\mathcal{P}(\mat{W}^{(n)})$ 
has a consistent directional derivative at $\mat{W}^{(n)}$ in certain direction with that of $\eqref{NLQCQP},$
we could show by using the results in \cite{2016-computing-B-stationary} \cite{2016-bolte-majorization} 
that $\mat{W}^{(\infty)}$ is a critical point to $\eqref{NLQCQP}$ under mild constraint qualification condition.

The DFGP method in \eqref{DFGP}  is an efficient matrix-free algorithm 
that are based solely on matrix-vector products and comparison
operations. The number of arithmetic operations per iteration for DFGP is $\mathcal{O}(K^2)$ or $\mathcal{O}(QMK),$
depending on the use of explicit or implicit matrix-vector multiplication $\mat{B}\vect{z}$.
At each step of the SCA algorithm, the number of additional
arithmetic operations for computing subgradients is $\mathcal{O}(QMK).$

\section{Simulation Results}\label{SR}
In this section, we provides numerical results to assess the performance of the proposed schemes, 
i.e., the SDR-G algorithm and the SCA-DFPG algorithm.  We assume that for each user, the small-scale fading 
is frequency-flat Rayleigh, i.e., complex Gaussian distributed with zero mean and unit variance, and the shadow fading 
is log-normally distributed with standard deviation $0.5$ dB.
For simplicity, we assume that all users have a common QoS target $\bar{\gamma}_q= 3$ dB
and the noise variance of each user is $\sigma_{k,q}^2=1 \forall k,q.$
The results are averaged over $500$ channel realizations.
The number of randomly generated candidates for each channel realization
is $L=1000$ and the number of iterations of DFGP method is set to $400.$

\subsection{Approximation Ratio Tests}
We first test the proposed SDR-G procedure listed in Algorithm 1 for homogeneous channel scenario
under various parameter settings. Tables 1 summarize the minimum value (Min), the maximum value (Max), 
the average value (Mean), and the standard deviation (Std) of empirical approximation 
ratios $v_{\text{SDR-G}}^*/v_{\text{SDR-LB}}$ over 500 independent channel realizations. 
We can see that the maximum values of $v_{\text{SDR-G}}^*/v_{\text{SDR-LB}}$
are lower than $3$ in all test examples. Moreover, the practical results are much better than those of
worst-case analysis. On the other hand, the minimum value, the maximum value and the average value 
of $v_{\text{SDR-G}}^*/v_{\text{SDR-LB}}$ all increase as $K$ grows for fixed $Q$ and $M$ in all test examples,
which also corroborates well with the theoretic analysis.
\begin{table}[!t] 
\centering
\caption{Statistics of empirical approximation ratios $v_{\text{SDR-G}}^*/v_{\text{SDR-LB}}$}
\begin{tabular}{|c|c|c||c|c|c|c||c|}
		\hline
		Q & M & K & Min & Max & Mean & Std & $\theta$ \\
		\hline
		2 & 8  & 10  & 1.0003 & 1.8816 & 1.4635 & 0.2527 & 15.81 \\
		2 & 8  & 20  & 1.2483 & 2.4106 & 1.8943 & 0.1996 & 22.36 \\
		2 & 8  & 30  & 1.5017 & 2.8103 & 2.2018 & 0.2179 & 27.39 \\
		2 & 16 & 10  & 1.0028 & 1.9346 & 1.4697 & 0.2642 & 15.81 \\
		2 & 16 & 20  & 1.3508 & 2.6042 & 1.9878 & 0.1989 & 22.36 \\
		2 & 16 & 30  & 1.7433 & 2.9791 & 2.3608 & 0.2170 & 27.39 \\
		\hline
		\hline
		3 & 8  & 10  & 1.0003 & 1.9114 & 1.4618 & 0.2496 & 10.77 \\
		3 & 8  & 20  & 1.2080 & 2.4799 & 1.8881 & 0.1958 & 13.57 \\
		3 & 8  & 30  & 1.6496 & 2.7887 & 2.2122 & 0.2260 & 15.54 \\
		3 & 16 & 10  & 1.0017 & 1.9406 & 1.4758 & 0.2655 & 10.77 \\
		3 & 16 & 20  & 1.3305 & 2.6277 & 1.9862 & 0.1963 & 13.57 \\
		3 & 16 & 30  & 1.7217 & 2.9683 & 2.3685 & 0.2163 & 15.54 \\
		\hline	
	\end{tabular}
\end{table}

\begin{figure}[!t]
	\centering
	\includegraphics[width=0.45\textwidth]{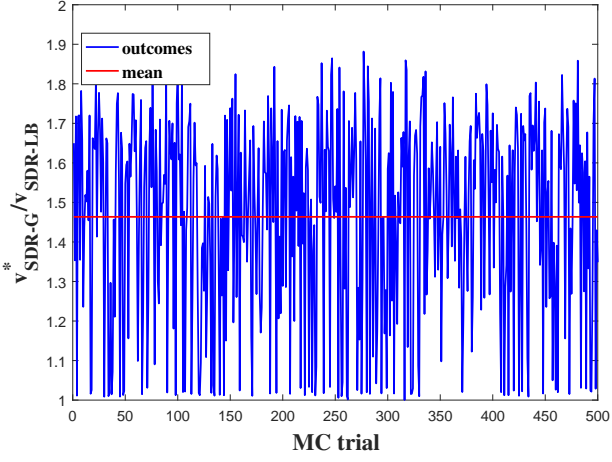}
	\caption{Empirical approximation ratios for $Q=2,M=8,K=10$.}
	\label{fig_UB}
\end{figure}

\begin{figure}[!t]
	\centering
	\includegraphics[width=0.45\textwidth]{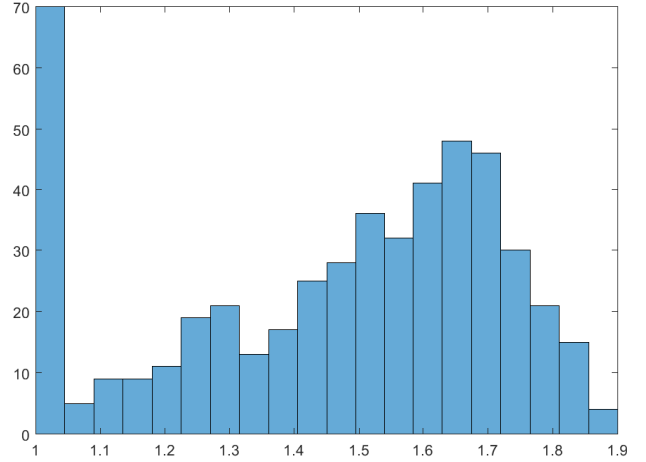}
	\caption{Histogram of the outcomes in Fig. 1.}
	\label{fig_UBHIST}
\end{figure}

Fig. \ref{fig_UB} plots the empirical approximation ratio of 500 independent channel realizations
for $Q=2,M=8,K=10.$ Fig. \ref{fig_UBHIST} shows the corresponding histogram. It can be seen that
in some cases the empirical approximation ratios are very near to $1,$ which means the optimal 
solutions are obtained by the proposed algorithm for these cases.

\subsection{Transmitting Power Comparisons}
\begin{figure}[!t]
	\centering
	\includegraphics[width=0.45\textwidth]{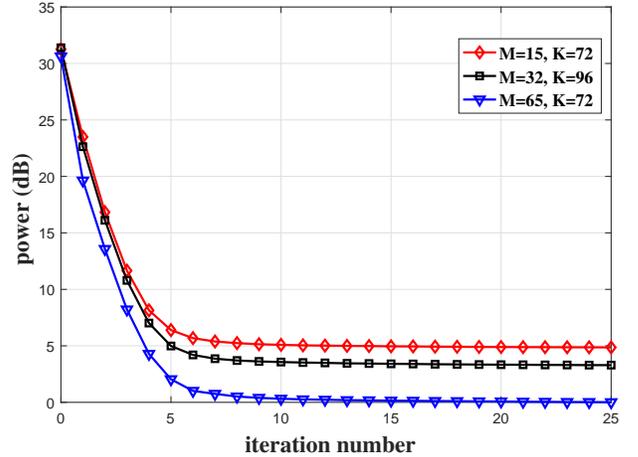}
	\caption{Convergence curve of the SCA-DFGP algorithm.}
	\label{fig_1}
\end{figure}

In this part, we  focus on general channel scenario and assume that there are $Q=3$ orthogonal channels. 
We first demonstrate the convergence of the SCA-DFGP algorithm. Fig.\ref{fig_1} 
plots the transmitting power consumption during each iteration for different settings in general channel scenario.
The results validate the monotonicity and convergence of the SCA-DFGP algorithm. 
It can be seen that at the first about $10$ iterations, the SCA-DFGP algorithm converges
very fast and reaches the major part of the limiting value.

We will next compare the the transmitting power consumption of the proposed co-design schemes 
with conventional scheduling algorithms. The first benchmark is fixed scheduling (OneGroup) \cite{2006TSP-SDR}, 
in which all users are scheduled into a single group and receive a common message in a fixed best channel. 
The second benchmark is random scheduling (Equipartition), in which all users 
are randomly scheduled into $Q$ groups with equal size. Moreover, the SDR lower bound (SDR-LB) is also presented.


\begin{figure}[!t]
	\centering
	\includegraphics[width=0.45\textwidth]{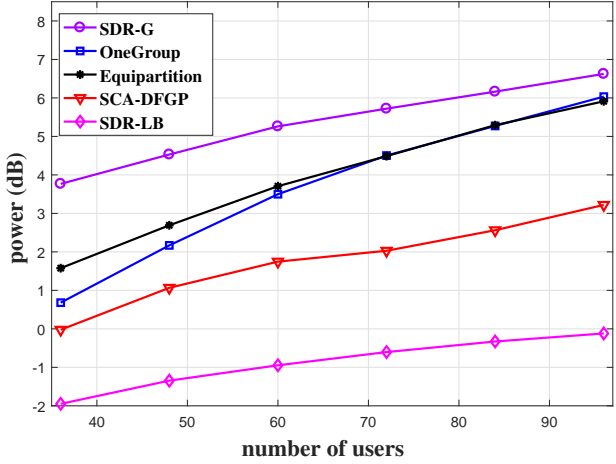}
	\caption{Transmitting power versus number of users, $K,$ for $M=32$ in general channel scenario.}
	\label{epsPowerVsK}
\end{figure}

\begin{figure}[!t]
	\centering
	\includegraphics[width=0.45\textwidth]{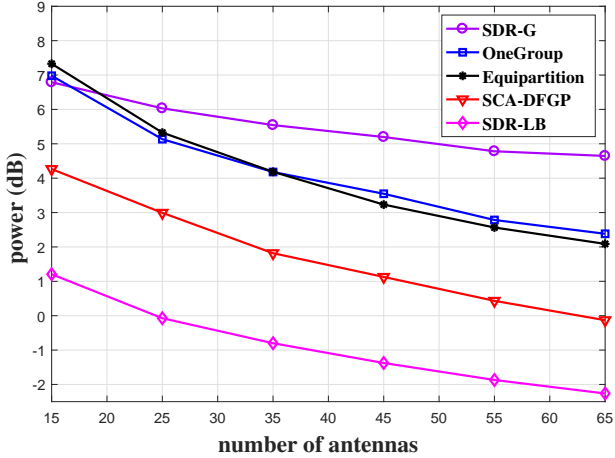}
	\caption{Transmitting power versus number of antennas, $M,$ for $K=72$ in general channel scenario.}
	\label{epsPowerVsM}
\end{figure}

Fig. \ref{epsPowerVsK} compares the average transmitting power of all the algorithms versus $K$ for $M = 32$.
Similarly, Fig. \ref{epsPowerVsM} compares the average transmitting power of all the algorithms versus $M$
for $K=72$. It can be seen from Fig. \ref{epsPowerVsK} and Fig. \ref{epsPowerVsM} that the average
transmitting power consumed by the SCA-DFGP algorithm is lower than the two benchmarks.
The performances of two benchmarks are very similar while the power saving of the SCA-DFGP algorithm 
over the two benchmarks is significantly beneficial especially when
the ratio of number of users to number of antennas is large.
On the other hand, the SDR-G algorithm performs poorly in  large-scale antenna arrays, 
especially when the number of users increases, which is also confirmed by the worst-case analysis and
many other studies \cite{2006TSP-SDR} \cite{2008TSP-MultiGroup} \cite{zhou2016fast}.
Moreover, the gap between the transmitting power for the SCA-DFGP algorithm and the SDR lower bound 
is always less than about $3$ dB in Fig. \ref{epsPowerVsK} and Fig. \ref{epsPowerVsM}.
Therefore, proper user scheduling is necessarily important when there are a large number of users in the system.


\section{Conclusions}\label{Conc}
In this paper, the joint multicast beamforming and user scheduling problem was investigated.
A mixed binary quadratically constrained quadratic program formulation and a highly-structured nonsmooth formulation
were presented. Convex-relaxation-based and convex-restriction-based algorithms were proposed to solve the problem.
Theoretical performance guarantee of convex-relaxation-based algorithm was proved and convergence of convex-restriction-based 
algorithm was established. Extensive numerical experiments were conducted to show the advantage of the proposed co-design
scheme over fixed scheduling and random scheduling in terms of power consumption.


\section{Appendix: Proof of Theorem 1}\label{sec14}

For any $\mat{W}^*\succeq 0$ and $\vect{\xi }\sim\mathcal{CN}(\vect{0},\mat{W}^*),$ it's easy to verify that 
$\abs{\vect{h}^\CT\vect{\xi }}^2$ is an exponential random variable with mean 
$\vect{h}^\CT\mat{W}^*\vect{h}$ and distribution function
\begin{equation}\label{exp_dist}
\Pr\left(\abs{\vect{h}^\CT\vect{\xi }}^2\le \eta\vect{h}^\CT\mat{W}^*\vect{h}\right)  = 1-e^{-\eta}\le \eta.
\end{equation}

Let independent random variables $\vect{\xi }_q\sim\mathcal{CN}(\vect{0},\mat{W}_q^*)\:(\forall q\in\mathcal{Q}).$ 
For any $\mu>0$ and $\eta>0$, we obtain
\begin{subequations}
\begin{align}
&\Pr\left(\sum_{q\in\mathcal{Q}}\norm{\vect{\xi }_q}^2\le \mu\sum_{q\in\mathcal{Q}}\Tr(\mat{W}_q^*),\min_{k \in \mathcal{K}} \max_{q\in\mathcal{Q}} \left\{\abs{\vect{h}_{k,q}^\CT\vect{\xi }_q}^2\right\} \ge \eta \right) \nonumber\\
 & \ge 1-\Pr\left(\sum_{q\in\mathcal{Q}}\norm{\vect{\xi }_q}^2\ge \mu\sum_{q\in\mathcal{Q}}\Tr(\mat{W}_q^*)\right) \nonumber\\
 &\quad-\sum_{k=1}^{K}\Pr\left(\max_{q\in\mathcal{Q}} \left\{\abs{\vect{h}_{k,q}^\CT\vect{\xi }_q}^2\right\} \le \eta\right) \\
& \ge 1-\frac{1}{\mu}-\sum_{k=1}^{K}\Pr\left(\max_{q\in\mathcal{Q}} \left\{\abs{\vect{h}_{k,q}^\CT\vect{\xi }_q}^2\right\} \le \eta\right),
\end{align}
\end{subequations}
where the first inequality is due to union bound of probability, and the last inequality is from the Markov's inequality. 

\subsection{General channel scenario}
From the constraints in \eqref{SDR1}, we have
\begin{equation}
\max_{q\in\mathcal{Q}} \vect{h}_{k,q}^\CT\mat{W}_q^*\vect{h}_{k,q} \ge \frac{1}{Q} \:\forall k\in \mathcal{K}.
\end{equation}
Let $q_k^*=\argmax_{q\in\mathcal{Q}} \vect{h}_{k,q}^\CT\mat{W}_q^*\vect{h}_{k,q}.$ For general channel scenario, we have
\begin{subequations}
\begin{align}
&\Pr\left(\max_{q\in\mathcal{Q}} \left\{\abs{\vect{h}_{k,q}^\CT\vect{\xi }_q}^2\right\} \le \eta\right) \nonumber\\
& \le \Pr\left(\max_{q\in\mathcal{Q}} \left\{\abs{\vect{h}_{k,q}^\CT\vect{\xi }_q}^2\right\} \le Q\eta\max_{q\in\mathcal{Q}} \vect{h}_{k,q}^\CT\mat{W}_q^*\vect{h}_{k,q}\right) \\
&=\Pi_{q\in\mathcal{Q}}\Pr\left(\abs{\vect{h}_{k,q}^\CT\vect{\xi }_q}^2\le Q\eta\max_{q\in\mathcal{Q}} \vect{h}_{k,q}^\CT\mat{W}_q^*\vect{h}_{k,q}\right) \label{indep}\\
&\le \Pr\left(\abs{\vect{h}_{k,q_k^*}^\CT\vect{\xi }_{q_k^*}}^2\le Q\eta \vect{h}_{k,q_k^*}^\CT\mat{W}_{q_k^*}^*\vect{h}_{k,q_k^*}\right) \\
& = Q\eta, \label{exp_eq}
\end{align}
\end{subequations}
where \eqref{indep} is due to independence of random variables $\{\vect{\xi }_q\}$ and \eqref{exp_eq} is from \eqref{exp_dist}.

Thus, by setting $\mu=\sqrt{5}$ and $\eta=\tfrac{1}{\sqrt{5}KQ},$ we have
\begin{subequations}
\begin{align}
&\Pr\left(\sum_{q\in\mathcal{Q}}\norm{\vect{\xi }_q}^2\le \mu\sum_{q\in\mathcal{Q}}\Tr(\mat{W}_q^*),\min_{k \in \mathcal{K}}\max_{q\in\mathcal{Q}} \left\{\abs{\vect{h}_{k,q}^\CT\vect{\xi }_q}^2\right\} \ge \eta \right) \nonumber \\
& \ge 1-\frac{1}{\mu}-KQ\eta \\
& = 1-\frac{2}{\sqrt{5}}=0.1056\dots.
\end{align}
\end{subequations}
We see that with positive probability of at least $0.1$, the randomly generated candidate beamformers 
 $\{\vect{x}_q^{(l)}\}$ satisfies
 \begin{equation}
\sum_{q\in\mathcal{Q}}\norm{\vect{x}_q^{(l)}}^2\le \sqrt{5}\sum_{q\in\mathcal{Q}}\Tr(\mat{W}_q^*)
\end{equation}	
and 
 \begin{equation}
\min_{k \in \mathcal{K}}\max_{q\in\mathcal{Q}}\left\{\abs{\vect{h}_{k,q}^\CT\vect{x}_q^{(l)}}^2\right\} \ge\tfrac{1}{\sqrt{5}KQ}.
\end{equation}
With $p(\{\vect{x}_q\})$ defined in \eqref{p_x}, $\left\{\sqrt{p(\{\vect{x}_q^{(l)}\})}\vect{x}_q^{(l)}\right\}$ is feasible for \eqref{MBQCQP},
so that
 \begin{subequations}\label{suc_prob}
\begin{align}
p^{(l)}&=p(\{\vect{x}_q^{(l)}\})\sum_{q\in\mathcal{Q}}\norm{\vect{x}_q^{(l)}}^2 \\
   &= \frac{\sum_{q\in\mathcal{Q}}\norm{\vect{x}_q^{(l)}}^2}{\min_{k \in \mathcal{K}}
   	\max_{q\in\mathcal{Q}}\left\{\abs{\vect{h}_{k,q}^\CT \vect{x}_q^{(l)}}^2\right\} } \\
   & \le \frac{\sqrt{5}\sum_{q\in\mathcal{Q}}\Tr(\mat{W}_q^*)}{1/(\sqrt{5}KQ)} \\
   & = 5KQ\cdot v_{\text{SDR-LB}}.
\end{align}
\end{subequations}
If one generates $L$ independent realizations of  $\{\vect{x}_q^{(l)}\}$ from $\mathcal{CN}(\vect{0},\mat{W}_q^*),$ it is
at least with probability $1-0.9^L$ to obtain one candidate beamformers satisfying \eqref{suc_prob}. Since 
$v_{\text{SDR-G}}^*=\min_{l=1,\dots,L}p^{(l)},$ it follows that 
 \begin{equation}
v_{\text{SDR-LB}} \le v^* \le v_{\text{SDR-G}}^* \le 5KQ\cdot v_{\text{SDR-LB}}.
\end{equation}
\subsection{Homogeneous channel scenario}
For homogeneous channel scenario, we have $\mat{W}_1^*=\mat{W}_2^*=\dots=\mat{W}_Q^*$,
$\vect{\tilde{h}}_{k}^\CT\mat{W}_1\vect{\tilde{h}}_{k} \ge 1 \: \forall k \in \mathcal{K}$
and $\vect{\xi }_q\sim\mathcal{CN}(\vect{0},\mat{W}_1^*)\:(\forall q\in\mathcal{Q}).$ 
Thus,
\begin{subequations}
	\begin{align}
	&\Pr\left(\max_{q\in\mathcal{Q}} \left\{\abs{\vect{\tilde{h}}_{k}^\CT\vect{\xi }_q}^2\right\} \le \eta\right) \nonumber\\
	& \le \Pr\left(\max_{q\in\mathcal{Q}} \left\{\abs{\vect{\tilde{h}}_{k}^\CT\vect{\xi }_q}^2\right\} \le \eta\vect{\tilde{h}}_{k}^\CT\mat{W}_1^*\vect{\tilde{h}}_{k}\right) \\
	&=\Pi_{q\in\mathcal{Q}} \Pr\left(\abs{\vect{\tilde{h}}_{k}^\CT\vect{\xi }_q}^2 \le \eta\vect{\tilde{h}}_{k}^\CT\mat{W}_1^*\vect{\tilde{h}}_{k}\right) \\
	&=\Pr\left(\abs{\vect{\tilde{h}}_{k}^\CT\vect{\xi }_1}^2 \le \eta\vect{\tilde{h}}_{k}^\CT\mat{W}_1^*\vect{\tilde{h}}_{k}\right)^Q \\
	&\le \eta^Q.
	\end{align}
\end{subequations}
By setting $\mu=\sqrt{5}$ and $\eta=\tfrac{1}{(\sqrt{5}K)^{(1/Q)}},$ we have
\begin{subequations}
	\begin{align}
	&\Pr\left(\sum_{q\in\mathcal{Q}}\norm{\vect{\xi }_q}^2\le \mu \Tr(\mat{W}_1^*),\min_{k \in \mathcal{K}}\max_{q\in\mathcal{Q}} \left\{\abs{\vect{\tilde{h}}_{k}^\CT\vect{\xi }_q}^2\right\} \ge \eta\right) \nonumber \\
	& \ge 1-\frac{1}{\mu}-K\eta^Q \\
	& = 1-\frac{2}{\sqrt{5}}=0.1056\dots.
	\end{align}
\end{subequations}
Similar to the proof for general channel scenario, we conclude that with probability of at least $1-0.9^L,$
if  $L$ independent realizations are generated, one could obtain an approximate solution such that  
\begin{equation}
v_{\text{SDR-LB}} \le v^* \le v_{\text{SDR-G}}^* \le 5K^{1/Q}\cdot v_{\text{SDR-LB}}.
\end{equation}


\begin{thebibliography}{1}
	
	\providecommand{\newblock}{\relax}
	\providecommand{\bibinfo}[2]{#2}
	\providecommand{\BIBentrySTDinterwordspacing}{\spaceskip=0pt\relax}
	\providecommand{\BIBentryALTinterwordstretchfactor}{4}
	\providecommand{\BIBentryALTinterwordspacing}{\spaceskip=\fontdimen2\font plus
		\BIBentryALTinterwordstretchfactor\fontdimen3\font minus
		\fontdimen4\font\relax}
	\providecommand{\BIBforeignlanguage}[2]{{%
			\expandafter\ifx\csname l@#1\endcsname\relax
			\typeout{** WARNING: IEEEtran.bst: No hyphenation pattern has been}%
			\typeout{** loaded for the language `#1'. Using the pattern for}%
			\typeout{** the default language instead.}%
			\else
			\language=\csname l@#1\endcsname
			\fi
			#2}}
	\providecommand{\BIBdecl}{\relax}
	\BIBdecl
	
\bibitem{2013SPM-MMIMO}
Rusek, F., Persson, D., Lau, B.K., Larsson, E.G., Marzetta, T.L., Edfors, O.,
et~al.: `Scaling up {MIMO}: Opportunities and challenges with very large
arrays', \emph{IEEE Signal Process Mag},  2013, \textbf{30}, (1), pp.~40--60

\bibitem{2012CM-eMBMS}
Lecompte, D., Gabin, F.: `Evolved multimedia broadcast/multicast service
({eMBMS}) in {LTE}-advanced: overview and {R}el-11 enhancements', \emph{IEEE
	Commun Mag},  2012, \textbf{50}, (11), pp.~68--74

\bibitem{2006TSP-SDR}
Sidiropoulos, N.D., Davidson, T.N., Luo, Z.Q.: `Transmit beamforming for
physical-layer multicasting', \emph{IEEE Trans Signal Process},  2006,
\textbf{54}, (6), pp.~2239--2251

\bibitem{2008TSP-MultiGroup}
Karipidis, E., Sidiropoulos, N.D., Luo, Z.Q.: `Quality of service and max-min
fair transmit beamforming to multiple cochannel multicast groups', \emph{IEEE
	Trans Signal Process},  2008, \textbf{56}, (3), pp.~1268--1279

\bibitem{2014-multicastPAPC}
Christopoulos, D., Chatzinotas, S., Ottersten, B.: `Weighted fair multicast
multigroup beamforming under per-antenna power constraints', \emph{IEEE Trans
	Signal Process},  2014, \textbf{62}, (19), pp.~5132--5142

\bibitem{2015-multicastPAPC}
Christopoulos, D., Chatzinotas, S., Ottersten, B.
\newblock `Multicast multigroup beamforming for per-antenna power constrained
large-scale arrays'.
\newblock In: Signal Process. Advances in Wireless Commun. (SPAWC), 2015 IEEE
16th International Workshop on. (IEEE,  2015. pp.~ 271--275

\bibitem{2013-coordinated}
Xiang, Z., Tao, M., Wang, X.: `Coordinated multicast beamforming in multicell
networks', \emph{IEEE Trans Wireless Commun},  2013, \textbf{12}, (1),
pp.~12--21

\bibitem{zhou2017coordinated}
Zhou, L., Zheng, L., Wang, X., Jiang, W., Luo, W.: `Coordinated multicell
multicast beamforming based on manifold optimization', \emph{IEEE Commun
	Lett},  2017, \textbf{21}, (7), pp.~1673--1676

\bibitem{2015-EnergyEfficientMulticast}
He, S., Huang, Y., Jin, S., Yang, L.: `Energy efficient coordinated beamforming
design in multi-cell multicast networks', \emph{IEEE Commun Lett},  2015,
\textbf{19}, (6), pp.~985--988

\bibitem{2015-coordinatedMulticastingUserSelection}
Hong, Y.W.P., Li, W.C., Chang, T.H., Lee, C.H.: `Coordinated multicasting with
opportunistic user selection in multicell wireless systems', \emph{IEEE Trans
	Signal Process},  2015, \textbf{63}, (13), pp.~3506--3521

\bibitem{zhou2016fast}
Zhou, L., Zhou, X., Chen, J., Jiang, W., Luo, W.
\newblock `Fast proximal gradient algorithm for single-group multicast
beamforming'.
\newblock In: Wireless Commun. \& Signal Process. (WCSP), 2016 8th Int. Conf.
on. (IEEE,  2016. pp.~ 1--5

\bibitem{matskani2009efficient}
Matskani, E., Sidiropoulos, N.D., Luo, Z.Q., Tassiulas, L.: `Efficient batch
and adaptive approximation algorithms for joint multicast beamforming and
admission control', \emph{IEEE Trans Signal Process},  2009, \textbf{57},
(12), pp.~4882--4894

\bibitem{shi2016smoothed}
Shi, Y., Cheng, J., Zhang, J., Bai, B., Chen, W., Letaief, K.B.: `Smoothed $
{L}_p $-minimization for green cloud-ran with user admission control',
\emph{IEEE J on Selected Areas in Commun},  2016, \textbf{34}, (4),
pp.~1022--1036

\bibitem{christopoulos2015multicast}
Christopoulos, D., Chatzinotas, S., Ottersten, B.: `Multicast multigroup
precoding and user scheduling for frame-based satellite communications',
\emph{IEEE Trans Wireless Commun},  2015, \textbf{14}, (9), pp.~4695--4707

\bibitem{zhou2015joint}
Zhou, H., Tao, M.
\newblock `Joint multicast beamforming and user grouping in massive {MIMO}
systems'.
\newblock In: Commun. (ICC), 2015 IEEE Int. Conf. on. (IEEE,  2015. pp.~
1770--1775

\bibitem{wu2012rank}
Wu, S.X., So, A.M., Ma, W.K.
\newblock `Rank-two transmit beamformed alamouti space-time coding for
physical-layer multicasting'.
\newblock In: Acoustics, Speech and Signal Process. (ICASSP), 2012 IEEE Int.
Conf. on. (IEEE,  2012. pp.~ 2793--2796

\bibitem{xu2014semidefinite}
Xu, Z., Hong, M., Luo, Z.Q.: `Semidefinite approximation for mixed binary
quadratically constrained quadratic programs', \emph{SIAM Journal on
	Optimization},  2014, \textbf{24}, (3), pp.~1265--1293

\bibitem{xu2016semidefinite}
Xu, Z., Hong, M.Y.: `Semidefinite relaxation for two mixed binary quadratically
constrained quadratic programs: Algorithms and approximation bounds', \emph{J
	of the Operations Research Society of China},  2016, \textbf{4}, (2),
pp.~205--221

\bibitem{zhang2013monotonic}
Zhang, Y.J.A., Qian, L., Huang, J., et~al.: `Monotonic optimization in
communication and networking systems', \emph{Foundations and Trends
	{\textregistered} in Networking},  2013, \textbf{7}, (1), pp.~1--75

\bibitem{luo2007approximation}
Luo, Z.Q., Sidiropoulos, N.D., Tseng, P., Zhang, S.: `Approximation bounds for
quadratic optimization with homogeneous quadratic constraints', \emph{SIAM
	Journal on optimization},  2007, \textbf{18}, (1), pp.~1--28

\bibitem{boyd2004convex}
Boyd, S., Vandenberghe, L.: `Convex optimization'.
\newblock (Cambridge university press,  2004)

\bibitem{2013-nesterov-introductory}
Nesterov, Y.: `Introductory lectures on convex optimization: A basic course'.
vol.~87.
\newblock (Springer Science \& Business Media,  2013)

\bibitem{2016-computing-B-stationary}
Pang, J.S., Razaviyayn, M., Alvarado, A.: `Computing {B}-stationary points of
nonsmooth {DC} programs', \emph{Mathematics of Operations Research},  2016,

\bibitem{2016-bolte-majorization}
Bolte, J., Pauwels, E.: `Majorization-minimization procedures and convergence
of {SQP} methods for semi-algebraic and tame programs', \emph{Mathematics of
	Operations Research},  2016, \textbf{41}, (2), pp.~442--465
\end{thebibliography}

\ifCLASSOPTIONcaptionsoff
\newpage
\fi

\end{document}